%% file: RoeverEtAl2014-preprint.tex
\documentclass[10pt,a4paper]{article}
\usepackage[authoryear,round]{natbib}
\usepackage{graphicx}
\usepackage[a4paper,hyperindex,
            colorlinks=true,linktocpage=false,
            bookmarks=false,
            breaklinks=true,
            pdftitle={Evidence synthesis for count distributions...},
            pdfauthor={Christian Roever, Stefan Andreas, Tim Friede},
            pdfkeywords={COPD; Meta analysis; Missing data; Negative binomial model.},
            linkcolor=black,citecolor=black,urlcolor=black]{hyperref}
\usepackage{breakurl}  
\usepackage{nicefrac}

\begin{document}
\title{Evidence synthesis for count distributions based on heterogeneous and incomplete aggregated data}
\author{Christian R\"{o}ver$^1$ \and Stefan Andreas$^{2,3}$ \and Tim Friede$^1$}

\footnotetext[1]{Department of Medical Statistics, University Medical Center G\"{o}ttingen, Humboldt\-allee~32, 37073~G\"{o}ttingen, Germany}
\footnotetext[2]{Lungenfachklinik Immenhausen, Robert-Koch-Stra{\ss}e~3, 34376~Immenhausen, Germany}
\footnotetext[3]{Clinics for Cardiology and Pulmonology, University Medical Center G\"{o}ttingen, Robert-Koch-Stra{\ss}e~40, 37099~G\"{o}ttingen, Germany}  

\maketitle

  \begin{abstract}
    \input{RoeverEtAl2014-abstract.tex}
  \end{abstract}

  \input{RoeverEtAl2014-main.tex}

  \bibliographystyle{plainnat}
  \bibliography{../../../literature/literature}

\end{document}

%% file: RoeverEtAl2014-abstract.tex
The analysis of count data is commonly done using Poisson models.
Negative binomial models are a straightforward and readily motivated
generalization for the case of overdispersed data, i.e., when the
observed variance is greater than expected under a Poissonian model.
%
%
%
Rate and overdispersion parameters then need to be considered jointly,
which in general is not trivial.
Here we are concerned with evidence synthesis in the case where the
reporting of data is rather heterogeneous, i.e., events are reported
either in terms of mean event counts, the proportion of event-free
patients, or rate estimates and standard errors.
Either figure carries some information about the relevant parameters,
and it is the joint modeling that allows for coherent inference on the
parameters of interest.
The methods are motivated and illustrated by a systematic review in
chronic obstructive pulmonary disease.

%% file: RoeverEtAl2014-main.tex
%
%

\providecommand{\prob}{\mathrm{P}}
\providecommand{\expect}{\mathrm{E}}
\providecommand{\var}{\mathrm{Var}}
\providecommand{\fevone}{$\mbox{FEV}_1$}
\providecommand{\normaldistn}{\mathrm{Normal}}
\providecommand{\unifdistn}{\mathrm{Unif}}
\providecommand{\tdistn}{\mathrm{t}}

\section{Introduction}
  Count data commonly occur as endpoints in clinical trials, for
  example, when one is interested in inferring rates of recurrent
  events.  
  Accordingly, these types of data are frequently encountered
  in meta analysis, when results from different studies
  are integrated.
%
%
  The Poisson distribution is
  often applicable when modeling event counts that are associated with
  a certain \textsl{rate} which then is usually the quantity of
  interest. The negative binomial distribution arises as a
  straightforward generalisation of the Poisson
  distribution \citep{Lawless1987}; it results as a marginal
  distribution of counts when the corresponding ``Poissonian'' rate is
  not fixed, but is associated with some uncertainty taking the
  mathematical form of a Gamma distribution. The introduced additional
  variability, the \textsl{overdispersion}, induces extra variation in
  the data and uncertainty in resulting estimates. The consideration
  of overdispersion is often useful and necessary in order to account
  for heterogeneity of some kind in the data.

  Clinical trials in the context of \textsl{chronic obstructive
  pulmonary disease (COPD)}, are commonly concerned with count
  data. The progress of COPD is characterized by
  recurrent \textsl{exacerbations}, periods of rapid worsening of the
  disease.  The presence of exacerbations, or their counts, are often
  used as clinical endpoints, as treatments are supposed to delay or
  prevent the occurence of exacerbations. Negative binomial models are
  commonly advocated and used in the context of COPD
  exacerbations \citep{KeeneEtAl2007,KeeneEtAl2008,AaronEtAl2008,EMA2012},
  but while the presence of overdispersion and the importance of its
  consideration appears to be undoubted, published evidence on its
  actual magnitude on the other hand is rather sparse. For
  example, \citet{AnzuetoEtAl2009} use ``\textsl{an over-dispersion
  estimate of~1.5}'' and \citet{CalverleyEtAl2009} use ``\textsl{a
  correction for overdispersion of~2}'', while the actual conventions
  used for quantifying overdispersion may be ambiguous.  Also, the way
  in which results are quantified varies considerably across studies;
  many studies quote exacerbation rate estimates, some provide
  standard errors (or confidence intervals) in addition, and some
  studies again quote numbers (or proportions) of patients with and
  without an exacerbation during the study.  

  Differences in the reporting of study results commonly pose
  problems when combining information sources that aggregate data not in
  the same way.  Off-the-shelf software may then not be suitable to deal with the peculiarities of a given
  problem.  The generic case of combining data of a common format will
  here be referred to as \textsl{meta analysis} while the more general case of
  combining data in different formats or possibly originating from different
  experimental designs is commonly called \textsl{evidence
  synthesis} \citep{SpiegelhalterEtAl}.
  Our aim was to set up a model to infer both unknowns while
  incorporating the different sources of data in a coherent manner.
  Here we aim for a model that consistently based on an
  underlying negative binomial process, to which the different data
  sources are linked in individual ways. The eventual formulation of a
  coherent data likelihood then allows for consistent inference based
  on all available information.

  In our case, both rate \textsl{and} overdispersion are important parameters
  determining the possible trial outcomes, while neither one is
  usually the figure of primary interest -- the actual focus usually
  lies on quantities like \textsl{rate ratios}.  
  If one knows the total number of exacerbations (or a rate estimate)
  and the number of zero-counts (exacerbation-free patients), one can
  immediately estimate rate and overdispersion jointly (e.g.\ via a
  maximum-likelihood approach). If only one of the figures is given,
  only one parameter may be estimated for the other being given or
  fixed. While we do not know any of the parameters precisely, we have
  some information on their probable order of magnitude from related
  studies. The idea now is to utilize a Bayesian
  framework \citep{SuttonAbrams2001,SpiegelhalterEtAl} to coherently
  incorporate the information from different sources. This way we are
  able incorporate exact likelihoods into the joint model, and to
  ``borrow strength'' among studies in order to make sense of the
  limited amount of information that may be contained in any single
  one.

  The outline of the paper is as follows. In
  Section~\ref{sec:copdIntro}, the context of count data in modeling
  exacerbation counts in COPD is
  introduced. Section~\ref{sec:modeling} describes Poisson and
  negative binomial models for count data in general and the problems
  arising with data encountered in studies investigating COPD\@. 
  The proposed approach to data modeling
  is developed, including a hierarchical model for the meta-analysis.
  In Section~\ref{sec:application}, the model is applied to actual
  data, and model variations are explored in order to investigate the
  impact of considering additional data.

\section{Motivating example: Meta-analysis in COPD}\label{sec:copdIntro}
  Chronic obstructive pulmonary disease (COPD) is a major cause of
  death and disability worldwide, and the burden of this disorder will
  continue to increase in the coming decades despite therapeutic
  advances \citep{DecramerJanssensMiravitlles2012}. Thus there is
  substantial need to improve symptomatic and prognostic burden in
  COPD\@.  Randomized controlled trials (RCTs) investigating a primary
  endpoint are the gold standard in proving efficacy of one therapy
  over another or over placebo.  Lung function such as forced
  expiratory volume in 1 second (\fevone) is a physiological measure
  and a global marker of disease severity in COPD\@. However,
  {\fevone} and other measures of lung function correlate poorly with
  patient-related outcomes such as symptoms, exercise tolerance,
  quality of life, exacerbations and mortality.  Mortality is a key
  endpoint, but two COPD trials failed to show significant effects on
  mortality, despite including about 6$\,$000 COPD patients each
  (TORCH \citep{CalverleyEtAl2007b}, UPLIFT \citep{TashkinEtAl2008c}).
  Exacerbations of COPD relate to mortality, impaired quality of life,
  lung function and health care costs.  Furthermore, the chronic and
  progressive course of COPD is frequently aggravated by
  exacerbations \citep{DecramerJanssensMiravitlles2012}.  Thus many
  phase~3 studies exploit COPD exacerbations as the primary
  endpoint \citep{CazzolaEtAl2008}.  There are different definitions
  of exacerbations mostly categorized by severity and the use of
  health care resources. Although all definitions harbour some
  subjective aspect, reproducibility and validity within one study is
  generally high. Problems arise, if different studies are accumulated
  and compared to each other as done in pooled- or
  meta-analyses 
  \citep{Suissa2006}, 
  which
  are usually addressed by the inclusion of variance components
  reflecting the heterogeneity in the statistical model.


  We are currently working on a meta-analysis comparing different
  treatments in COPD (R\"{o}ver et~al., in preparation).  In order to
  find relevant studies, we performed a literature search using
  PubMed; in addition we considered studies that were cited in the
  meta-analyses
  by \citet{PuhanEtAl2009,MillsEtAl2011,IQWIG2012,KarnerChongPoole2012,DongEtAl2012}.
  In the present analysis, we will focus on the subset of studies
  comparing \textsl{long-acting muscarinic antagonist (LAMA)} with
  placebo treatment.

\section{Statistical model}\label{sec:modeling}
\subsection{Notation and basic properties}\label{sec:poissonNB}
  A Poisson model naturally arises as a model for event counts that
  are associated with a certain
  rate \citep[Ch.~2.6]{SpiegelhalterEtAl}. The Poisson model is
  specified through a single parameter, the \textsl{rate}~$\lambda>0$,
  which is usually to be interpreted \textsl{per unit time}, i.e., as
  a \textsl{yearly} rate, for example. The number of events occuring
  over a duration $\delta$ then follows a Poisson distribution with
  mean~$\delta\lambda$; the expected number as well as its variance
  are equal to $\delta\lambda$.  The negative binomial distribution in
  addition possesses an overdispersion parameter~$\varphi\geq 0$. For
  $\varphi=0$, the model again simplifies to the Poissonian. With
  $\varphi>0$ the distribution's expectation remains the same, but the
  variance increases to $\delta\lambda
  (1+\varphi \delta\lambda)$ \citep{Lawless1987,Hilbe2011}. The negative
  binomial model arises as a Poisson mixture distribution, where,
  instead of being constant, the rate is a Gamma distributed random
  variable with expectation~$\lambda$ and variance~$\lambda^2\varphi$,
  so that the rate's coefficient of variation is $\sqrt{\varphi}$;
  this way the negative binomial model is able to account for
  extra-Poissonian heterogeneity in the data.

\subsection{Illustrating example}
  \begin{figure}[t]
    \begin{center}
      \includegraphics[width=0.66\linewidth]{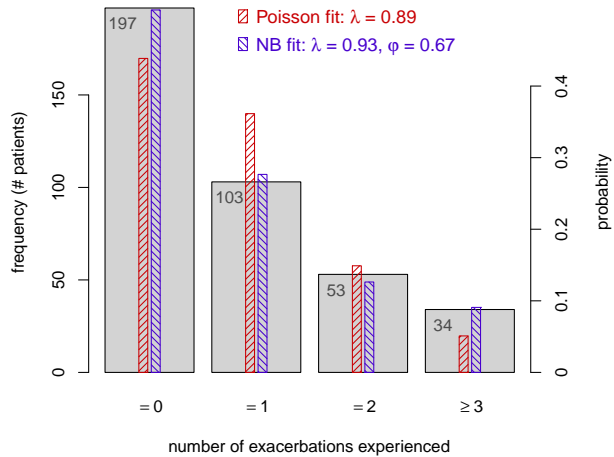}
      \caption{\label{fig:SethiData}An example of actual exacerbation
        counts observed in placebo-treated patients over a duration of
        48~weeks taken from \citet{SethiEtAl2010}. A negative binomial
        model fits the data better than a Poisson model as it allows
        for a greater fraction of extreme (i.e., zero or large)
        counts, and accordingly also fits better in the moderate
        range.
        95\% confidence intervals for the Poisson and negative binomial 
        rates are [0.79, 0.99] and [0.80,1.07], respectively. 
        The overdispersion's 95\% confidence interval is [0.30, 1.04].}  
    \end{center} 
  \end{figure}
  Figure~\ref{fig:SethiData} provides an illustration of the
  difference between Poisson and negative binomial models in
  comparison with actual count data. The bars represent the observed
  frequencies of exacerbation counts in an actual
  study \citep{SethiEtAl2010}.  The red and blue lines show the
  best-fitting Poisson and negative binomial models (fitted via
  maximum likelihood) for the given data. One can see that while the
  two rate parameters barely differ, the additional consideration of a
  non-zero overdispersion provides a much better fit to the data, as
  it allows to account for the increased number of ``extreme''
  outcomes (zero or large exacerbation counts) that would be unlikely
  to occur under a plain Poissonian model. In the negative binomial
  model, the standard error associated with the rate estimate is
  larger than in the Poisson model 
  by a factor of a third
  (0.068 instead of 0.051).  Looking at the data, one can already see
  that the two commonly quoted summary statistics of (mean) rate and
  the number of exacerbation-free patients carry somewhat
  complementary pieces of information which in combination allow to
  infer both $\lambda$ and $\varphi$.

\subsection{Parameter estimation and sufficient statistics}
  When we assume a Poisson model for the observed event counts~$x_i$
  in a group of $n$ patients that were all observed over
  durations~$\delta_i$, then all that is required in order to infer
  the unkown rate~$\lambda$ is the total count $t=x_1+\ldots+x_n$ (and
  the total observation time $\delta_t=\delta_1+\ldots+\delta_n$), as
  the total constitutes the \textsl{sufficient statistic} in this
  case.  These numbers are rarely explicitly quoted, but being
  provided with a rate estimate $\hat{\lambda}=\nicefrac{t}{\delta_t}$
  (which is both the maximum-likelihood and moment estimator) and
  assuming a constant observation duration
  ($\delta_1=\ldots=\delta_n=\delta$), one can again directly infer
  the total~$t$.  In the case of a negative binomial distribution on
  the other hand, there is no simple form for a sufficient statistic;
  ideally one would need the complete data (all individual
  counts~$x_i$) in order to derive for example maximum-likelihood
  estimates \citep{Lawless1987}. Only for the case of constant
  durations~$\delta_i=\delta$, the maximum-likelihood estimator for
  the rate is the same as in the Poisson case, and there is also a
  moment estimator for the overdispersion available, although its
  properties are questionable, as it may also turn out negative.

  Another common alternative approach to modeling count data of the
  present kind is to consider \textsl{odds ratios}, i.e., to compare
  chances of an exacerbation (at least one event vs.\ no event)
  between treatment groups; this approach is used e.g.\
  by \citet{PuhanEtAl2009,IQWIG2012,KarnerChongPoole2012}. However, if
  we assume an underlying Poisson or negative binomial process, the
  odds ratio only equals the rate ratio in the limit of small rates or
  small exposure durations. Otherwise odds ratios and rate ratios will
  in general differ, depending on the exposure duration and the amount
  of overdispersion present, and hence are not directly comparable
  (see Appendix~A.1 for an explicit derivation).

\subsection{Available data}
  In the studies that we will investigate in the following, the most
  commonly used figures for quantifying disease severity are the
  number of exacerbation-free patients, rate estimates, partly in
  conjunction with confidence intervals or standard errors, or the
  median time until the first exacerbation. Standard errors and
  confidence intervals can easily be converted into one
  another \citep{HigginsGreen}, and numbers of exacerbation-free
  patients are commonly derived from the total number of patients and
  the provided percentage. For a given rate estimate $\hat{\lambda}$,
  we assume that it results as the total count~$t$ divided by the
  cumulative observation time~$nd$ in order to infer the total number
  as $t=\hat{\lambda}n\delta$.

\subsection{Marginal and joint modeling of total counts and zero-counts\label{sec:TZcounts}}
  Suppose that a total of $n$~patients are observed in a trial of duration
  $\delta$. Each patient's event count follows a negative
  binomial distribution with rate~$\delta \lambda$ and
  overdispersion~$\varphi$.
  Let $T$ be the total number of events observed; by the central limit
  theorem its marginal distribution is approximately normal with
  moments
  \begin{equation}
    \expect[T] \;=\; n \delta \lambda \qquad \mbox{and} \qquad
    \var(T)    \;=\; n \delta \lambda \, (1+\varphi \delta \lambda). \label{eqn:NBMoments}
  \end{equation}

  The marginal distribution of the number of ``zero-counts'', the
  number of event-free patients, $Z$, is Binomial with size~$n$ and
  probability~$\pi_0$ given by the probability of a zero outcome for
  the negative binomial model,
  \begin{equation}
    \pi_0 \;=\;
    \left\{\begin{array}{ll}
          {\textstyle (1+\varphi \delta \lambda)^{-\frac{1}{\varphi}}} & \mbox{for } \varphi>0 \\
          {\textstyle \exp(-\delta\lambda)} & \mbox{for } \varphi=0.
          \end{array}\right.
  \end{equation}


  $T$ and $Z$ are not stochastically independent, rather they are
  negatively correlated; i.e., when there is a large number of
  zeroes in the data, we expect a small total count.  An
  approximate joint probability distribution of $T$ and $Z$ can be
  derived by extending the ideas above. We can split the joint
  probability density into the factors $p(t,z)=p(z)\times p(t|z)$. The
  conditional density $p(t|z)$ describes the probability distribution
  of the total count~$T$ for which we know the number~$Z\!=\!z$ of zeroes in
  the total sum of negative binomial samples. The total $T$ thus is
  constituted of $n\!-\!z$ summands that are drawn from a ``truncated''
  negative binomial distribution excluding the zero outcome. We can
  again determine the first two moments of the truncated distribution
  and use the normal approximation as above; see Appendix~A.2 for the
  explicit derivation.  The marginal and joint distributions of $T$
  and $Z$ then define the likelihood function for the data that is
  used in the following.

\subsection{The hierarchical model}\label{sec:model}
  In the following we set up a hierarchical model with study-level
  rate and overdispersion parameters, allowing for heterogeneity
  between studies in both parameters. Depending on the particular type
  of data provided, different studies then contribute different pieces
  of information on both parameters.

  We have data on $k$ studies investigating both treatment and
  placebo.  Each study~$i$ has an associated study duration~$\delta_i$
  and is comprised of $\ell_i$ study arms.  The $i$th study's $j$th arm
  corresponds to a treatment category indexed by~$m_{ij}$
  ($m_{ij}\in\{0,1\}$, where $m_{ij}=0$ indicates a placebo group),
  and has a number~$n_{ij}$ of patients associated. Some study arms
  provide a rate estimate~$\hat{\lambda}_{ij}$ together with an
  associated standard error, some provide only their total
  exacerbation count $T_{ij}$, (corresponding to $\hat{\lambda}_{ij}
  n_{ij} \delta_i$ for some rate estimate $\hat{\lambda}_{ij}$), some
  provide the number $Z_{ij}$ of zero-counts (exacerbation-free
  patients), and some provide both $T_{ij}$ and $Z_{ij}$.  Studies
  directly providing rate estimates and standard errors are considered
  via the common normal approximation to the
  likelihood \citep{SpiegelhalterEtAl}. Studies quoting one or both of
  $T$ and $Z$ are considered in the analysis via the likelihood
  function described in Section~\ref{sec:TZcounts}.
  While one might use the information on rates and standard errors
  within a negative binomial model (which might in fact be considered
  an overall more consistent approach), here we used the normal
  approximation in order to demonstrate the gain compared to the most
  basic ``classical'' approach.

  The hierarchical model is parameterized as follows.  The rate for
  the $j$th arm of the $i$th study is simply given by $\lambda_i$ if
  it is a placebo group (i.e., if $m_{ij}=0$), or otherwise by
  $\lambda_i\times\vartheta\times\psi_{ij}$.  The parameter of primary
  interest is $\vartheta$, the treatment effect, and $\psi_{ij}$ is a
  random effect accounting for particularities in treatment, dosing,
  etc.\ differing between studies or study arms. The overdispersion in
  the $i$th study is given by $\varphi_i$.

  The prior for each trial's ``placebo'' rate~$\lambda_i$ is normal on
  the log-scale (i.e., log-normal):
  \begin{equation}
    \log(\lambda_i)  \;\sim\;  \normaldistn\bigl(\mu_\lambda, \sigma^2_\lambda\bigr). \nonumber
  \end{equation}
  The corresponding hyperparameters 
  ($\mu_\lambda$, $\sigma^2_\lambda$) 
  are assigned vague uniform prior distributions on the log scale:
  \begin{equation}
    \mu_\lambda  \;\sim\;  \unifdistn\bigl(\textstyle\log(10^{-2}), \, \log(10^2)\bigr),
     \quad
    \log(\sigma_\lambda)  \;\sim\;  \unifdistn\bigl(\textstyle\log(10^{-3}), \, \log(10)\bigr).
     \nonumber
  \end{equation}
  Similarly, each trial's overdispersion~$\varphi_i$ also has a normal prior on
  the log-scale, i.e.,
  \begin{equation}
    \log(\varphi_i)  \;\sim\;  \normaldistn\bigl(\mu_\varphi, \sigma^2_\varphi\bigr), \nonumber
  \end{equation}
  again with similar priors for the hyperparameters ($\mu_\varphi$,
  $\sigma^2_\varphi$):
  \begin{equation}
    \mu_\varphi  \;\sim\;  \unifdistn\bigl(\textstyle\log(10^{-4}), \, \log(10^4)\bigr),
     \quad
    \log(\sigma_\varphi)  \;\sim\;  \unifdistn\bigl(\textstyle\log(10^{-3}), \, \log(10)\bigr).
     \nonumber
  \end{equation}
  For the treatment effect we use a normal prior that is centered
  around a neutral effect (i.e., $\vartheta=1$) and loosely confined
  to within the range $[\nicefrac{1}{10},10]$ with $\approx 90\%$
  probability, i.e.
  \begin{equation}
    \log(\vartheta) \;\sim\; \normaldistn\bigl(0,\,\log(4)^2\bigr). \nonumber
  \end{equation}
  For each arm's random effect we again use a vague and also
  heavy-tailed Student\mbox{-}$t$-prior with 3~degrees of freedom
  centered around $\psi_{ij}=1$, i.e., no effect: 
  \begin{equation}
    \log(\psi_{ij}) \;\sim\; \tdistn_3\bigl(0,\,\sigma^2_\psi \bigr). \nonumber
  \end{equation}
  The heavy-tailed prior here is meant to allow for individual odd
  studies and to bound their influence on the overall result.  The
  scale parameter $\sigma_\psi$ again is an unknown with a uniform
  prior on the log scale:
  \begin{equation}
    \log(\sigma_\psi) \;\sim\; 
    \unifdistn\bigl(\textstyle \log(10^{-3}),\, \log(10) \bigr). \nonumber
  \end{equation}
  The parameters accounting for heterogeneity ($\sigma_\lambda$,
  $\sigma_\varphi$ and $\sigma_\psi$) here are modelled on their
  logarithmic scales. In a related setting, similar priors (uniform
  for the logarithmic heterogeneity) have been investigated among a
  range of other uninformative or weakly informative choices, which
  were all found to yield comparable inferences as long as the number
  of considered studies was not too low \citep{LambertEtAl2005}.  The
  priors used here are truncated versions of the (otherwise improper)
  Jeffreys priors for location and scale
  parameters \citep{Jaynes1968,Jeffreys1946}.
  The study-specific rate and overdispersion parameters ($\lambda_i$
  and $\phi_i$) and the arm-specific random effect ($\psi_{ij}$) are
  for now treated as a~priori independent.

\section{Application to COPD example}\label{sec:application}
\subsection{Investigated studies}\label{sec:SysLitRev}
  Our literature search resulted in 24~placebo-controlled studies
  investigating LAMA treatment and providing information on
  exacerbations.
  The resulting numbers of studies providing total
  exacerbation counts and/or the number of zero-counts
  (exacerbation-free patients) are shown in
  Table~\ref{tab:studyCounts}.
  \begin{table}[t]
  \begin{center}
  \caption{\label{tab:studyCounts}Four of the studies found provide rate
    estimates along with standard errors. The below table shows the
    numbers of studies \textsl{without standard errors}, but providing
    a total count (or rate) or the number of zero-counts. Another
    8~studies provide both, and also considering studies only giving one
    of the two, we can include another 12~additional studies in the
    analysis.}
  \begin{tabular}{cccc}
  \hline \hline
  &           & \multicolumn{2}{c}{\textsl{total count provided?}}\\
  &           & No & Yes\\
  \hline
  \textsl{zero count} & No & 0 & 3\\
  \textsl{provided?}  & Yes & 9 & 8\\
  \hline\hline
  \end{tabular}
  \end{center}
  \end{table}
  Based on the different types of information provided, we defined 3
  subsets of the data, namely: 
  (A)~the studies providing rate estimates and standard errors,
  (B)~the above studies, plus the ones giving both total counts \textsl{and} zero-counts, and
  (C)~the above studies, plus the ones giving either total counts \textsl{or} zero-counts.
  The (overlapping) data subsets (A), (B) and (C) consist of 4, 12,
  and 24 studies, respectively.  
  \nocite{BatemanEtAl2010b,DonaldsonEtAl2012,PowrieEtAl2007,TashkinEtAl2008c}
  All data are shown in Table~\ref{tab:data}.

  \begin{table}[t]
  \centering 
  \caption{\label{tab:data}The example data underlying the analyses. Each study comprises a placebo arm (P) and one or two LAMA treatment arms (L)\@.
Some studies report exacerbation rates along with standard errors (group~A). 
Other studies in addition report both the number of exacerbation-free patients (zero-counts) and the total exacerbation count (group~B), and yet a larger set reports at least one of the two (group~C).
When rates are quoted, the total may be derived from the study duration and number of included patients.}
  \footnotesize
  \setlength{\tabcolsep}{0.8ex}
  \begin{tabular}{lllrrccrr}
    \hline
\hline
first author (year) & group & arm & patients & duration (yr) & rate & std.\ err. & total & zeroes \\ 
  \hline
Ambrosino (2008) & B, C & P & 106 & 0.4808 &  &  & 26 & 85 \\ 
   &  & L & 103 & 0.4808 &  &  & 19 & 89 \\ 
  Bateman (2010a) & B, C & P & 653 & 1.0000 & 1.91 &  & 1247 & 365 \\ 
   &  & L & 670 & 1.0000 & 0.93 &  & 623 & 421 \\ 
   &  & L & 667 & 1.0000 & 1.02 &  & 680 & 421 \\ 
  Bateman (2010b) & A, B, C & P & 1953 & 0.9231 & 0.87 & 0.051 & 1568 & 1112 \\ 
   &  & L & 1939 & 0.9231 & 0.69 & 0.0561 & 1235 & 1256 \\ 
  Beeh (2006) & C & P & 403 & 0.2308 &  &  &  & 323 \\ 
   &  & L & 1236 & 0.2308 &  &  &  & 1056 \\ 
  Brusasco (2003) & C & P & 400 & 0.5000 & 1.49 &  & 298 &  \\ 
   &  & L & 402 & 0.5000 & 1.07 &  & 215 &  \\ 
  Casaburi (2002) & B, C & P & 371 & 1.0000 & 0.95 &  & 352 & 215 \\ 
   &  & L & 550 & 1.0000 & 0.76 &  & 418 & 352 \\ 
  Chan (2007) & B, C & P & 305 & 0.9230 & 0.92 &  & 259 & 180 \\ 
   &  & L & 608 & 0.9230 & 0.88 &  & 494 & 340 \\ 
  Cooper (2010) & C & P & 225 & 1.8460 & 0.54 &  & 194 &  \\ 
   &  & L & 239 & 1.8460 & 0.51 &  & 208 &  \\ 
  Donaldson (2012) & A, B, C & P & 52 & 1.0000 & 2.129 & 0.3206 & 111 &  \\ 
   &  & L & 48 & 1.0000 & 0.9855 & 0.2283 & 47 &  \\ 
  Donohue (2002) & C & P & 201 & 0.5000 &  &  &  & 109 \\ 
   &  & L & 209 & 0.5000 &  &  &  & 132 \\ 
  Donohue (2010) & C & P & 418 & 0.5000 & 0.72 &  & 150 &  \\ 
   &  & L & 415 & 0.5000 & 0.53 &  & 110 &  \\ 
  D'Urzo (2011) & C & P & 260 & 0.5000 &  &  &  & 197 \\ 
   &  & L & 532 & 0.5000 &  &  &  & 439 \\ 
  Dusser (2006) & B, C & P & 510 & 1.0000 & 1.69 &  & 862 & 238 \\ 
   &  & L & 500 & 1.0000 & 1.1 &  & 550 & 238 \\ 
  Freeman (2007) & B, C & P & 162 & 0.2308 &  &  & 44 & 160 \\ 
   &  & L & 182 & 0.2308 &  &  & 23 & 181 \\ 
  Johansson (2008) & C & P & 117 & 0.2308 &  &  &  & 113 \\ 
   &  & L & 107 & 0.2308 &  &  &  & 105 \\ 
  Moita (2008) & C & P & 160 & 0.2500 &  &  &  & 156 \\ 
   &  & L & 144 & 0.2500 &  &  &  & 140 \\ 
  Niewoehner (2005) & B, C & P & 915 & 0.5000 & 1.05 &  & 480 & 620 \\ 
   &  & L & 914 & 0.5000 & 0.85 &  & 388 & 659 \\ 
  Powrie (2007) & A, B, C & P & 73 & 1.0000 & 2.46 & 0.4471 & 134 & 26 \\ 
   &  & L & 69 & 1.0000 & 1.17 & 0.2709 & 60 & 39 \\ 
  Rio (2007) & C & P & 125 & 0.2308 &  &  &  & 116 \\ 
   &  & L & 123 & 0.2308 &  &  &  & 113 \\ 
  Tashkin (2008) & A, B, C & P & 3006 & 4.0000 & 0.85 & 0.02 & 10220 & 955 \\ 
   &  & L & 2986 & 4.0000 & 0.73 & 0.02 & 8719 & 986 \\ 
  Tonnel (2008) & B, C & P & 288 & 0.7500 & 1.83 &  & 395 & 158 \\ 
   &  & L & 266 & 0.7500 & 1.05 &  & 209 & 165 \\ 
  Troosters (2011) & C & P & 219 & 0.2308 &  &  &  & 194 \\ 
   &  & L & 238 & 0.2308 &  &  &  & 228 \\ 
  Verkindre (2006) & C & P & 46 & 0.2308 &  &  &  & 43 \\ 
   &  & L & 45 & 0.2308 &  &  &  & 45 \\ 
  Voshaar (2008) & C & P & 181 & 0.2308 &  &  &  & 160 \\ 
   &  & L & 180 & 0.2308 &  &  &  & 155 \\ 
   &  & L & 180 & 0.2308 &  &  &  & 164 \\ 
   \hline
\hline
\end{tabular}
\end{table}

\subsection{Implementation}
  We utilized the \textsl{JAGS} software \citep{Plummer2003}\nocite{jags340}
  in conjunction with the ``rjags'' package \citep{rjags} in order to
  carry out stochastic (Monte Carlo) integration of the unknown
  parameters' posterior distribution \citep{BDA}.
  In order to ensure convergence of the algorithm, we ran four MCMC chains in parallel, starting from different overdispersed initial values.
  Chains were run for $10^7$~iterations, where the first 10\%~of samples were discarded as burn-in. Correlation between subsequent samples was reduced by using only every 100th sample.
%
%
  Convergence of parallel chains to the same mode was checked using
  the (multivariate) potential scale reduction
  factor \citep{BrooksGelman1998}, which was well below $0.1\%$ in all
  cases.
  In the following, we show estimates,
  marginal posterior probability density functions etc.\ based on
  these Monte Carlo samples. The densities shown in
  figures~\ref{fig:thetaPosterior} and~\ref{fig:hyperParams} are based
  on kernel density estimates derived from the samples.

\subsection{Results}\label{sec:results}
\begin{figure}[t]
    \begin{center}
      \includegraphics[width=0.45\linewidth]{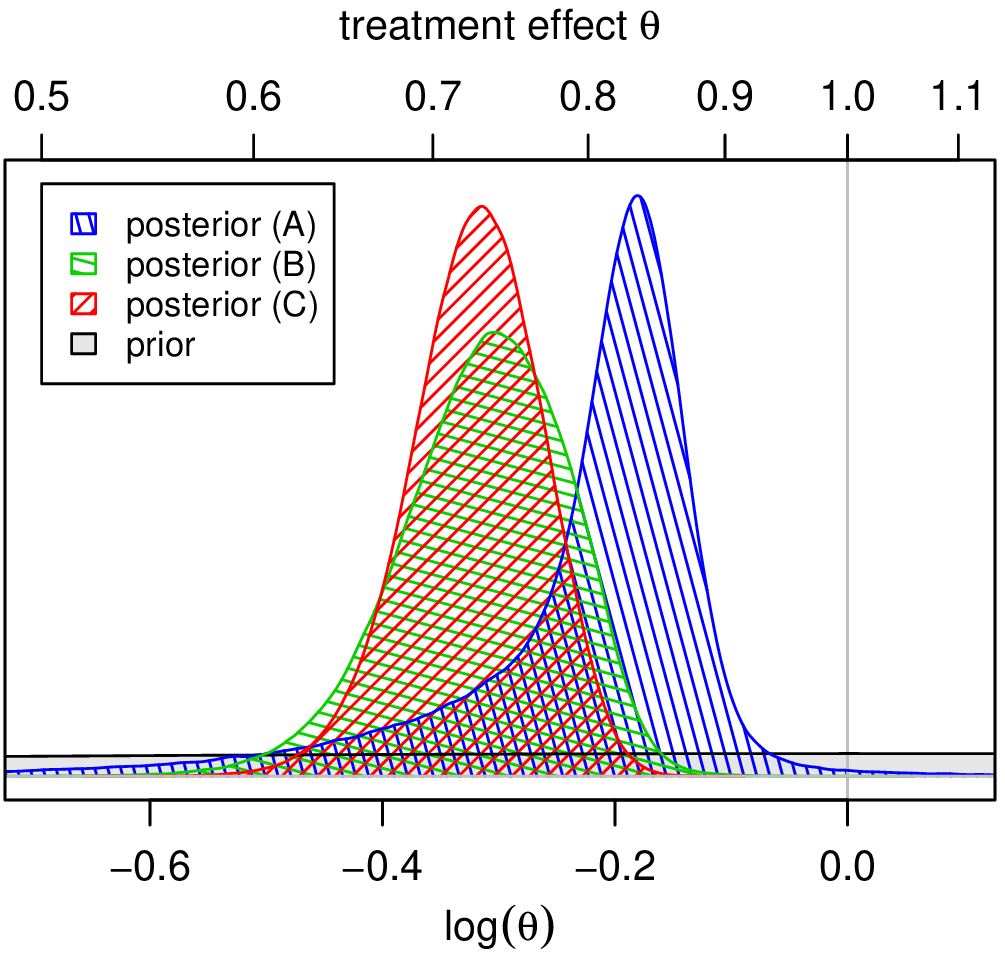}
      \hspace{0.05\linewidth}
      \includegraphics[width=0.45\linewidth]{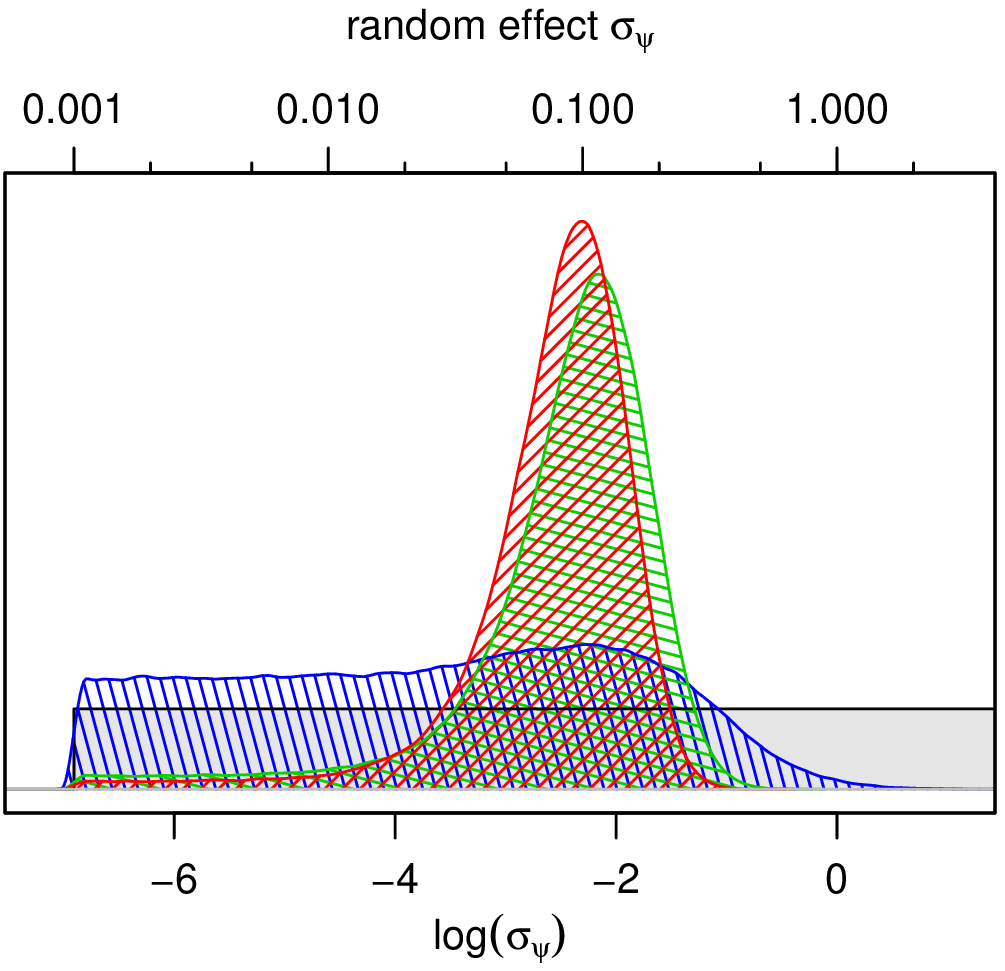}
      \caption{\label{fig:thetaPosterior}Marginal posterior
        probability density functions (kernel density estimates)
        based on different amounts of data. 
        The prior density is also shown in comparison.}
    \end{center}
  \end{figure}

  We applied the above model to the complete data, and its subsets as
  described in Section~\ref{sec:SysLitRev}.
  Figure~\ref{fig:thetaPosterior} shows the marginal posterior
  distribution of the parameter of primary interest, the treatment
  effect~$\vartheta$ for all three scenarios.
  One can see that the posterior distributions based on different data
  subsets are consistent with each other, and that with the inclusion
  of additional studies the evidence in favour of a substantial
  treatment effect is increased.

  While only approximately comparable (see
  Section~\ref{sec:poissonNB}), the effect on the annualized
  exacerbation rate is of the same order of magnitude as the odds
  ratio estimates found in related meta-analyses
  (\citet{PuhanEtAl2009}: 0.71, \citet{IQWIG2012}:
  0.76, \citet{KarnerChongPoole2012}: 0.78).  Performing a simple
  random-effects meta-analysis \citep{Viechtbauer2010} 
  on the (logarithmic) rate ratios of
  subset~(A) 
  yields a combined
  estimate for the treatment effect of 0.69 and a 95\% confidence
  interval of [0.52, 0.93], which is in line with the results from the
  Bayesian analysis 
  yielding a median of~0.82
  and credibility interval [0.56, 0.92].
  Interestingly, the estimates 
  differ by half a standard error, while the confidence /
  credibility limits still coincide.  With the consideration of 8
  additional studies reporting both a rate estimate and the proportion
  of event-free patients, the estimate changes to 0.73 [0.63, 0.83],
  and the inclusion of 12~studies providing one of the two figures
  narrows the estimate further down 
  to 0.73 [0.65, 0.80].

  The posterior distribution of the heterogeneity parameter
  $\sigma_\psi$ (right panel of Figure~\ref{fig:thetaPosterior}) shows
  that the observed data allow to constrain the heterogeneity's
  magnitude to an order of $\approx 0.1$ or below.  While using only
  data subset (A) we do not gain much information beyond that on this
  parameter, the inclusion of additional data allows to constrain the
  heterogeneity to be of a magnitude larger than, say $\approx 0.01$.

\begin{figure}[t]
    \begin{center}
      \includegraphics[width=0.30\linewidth]{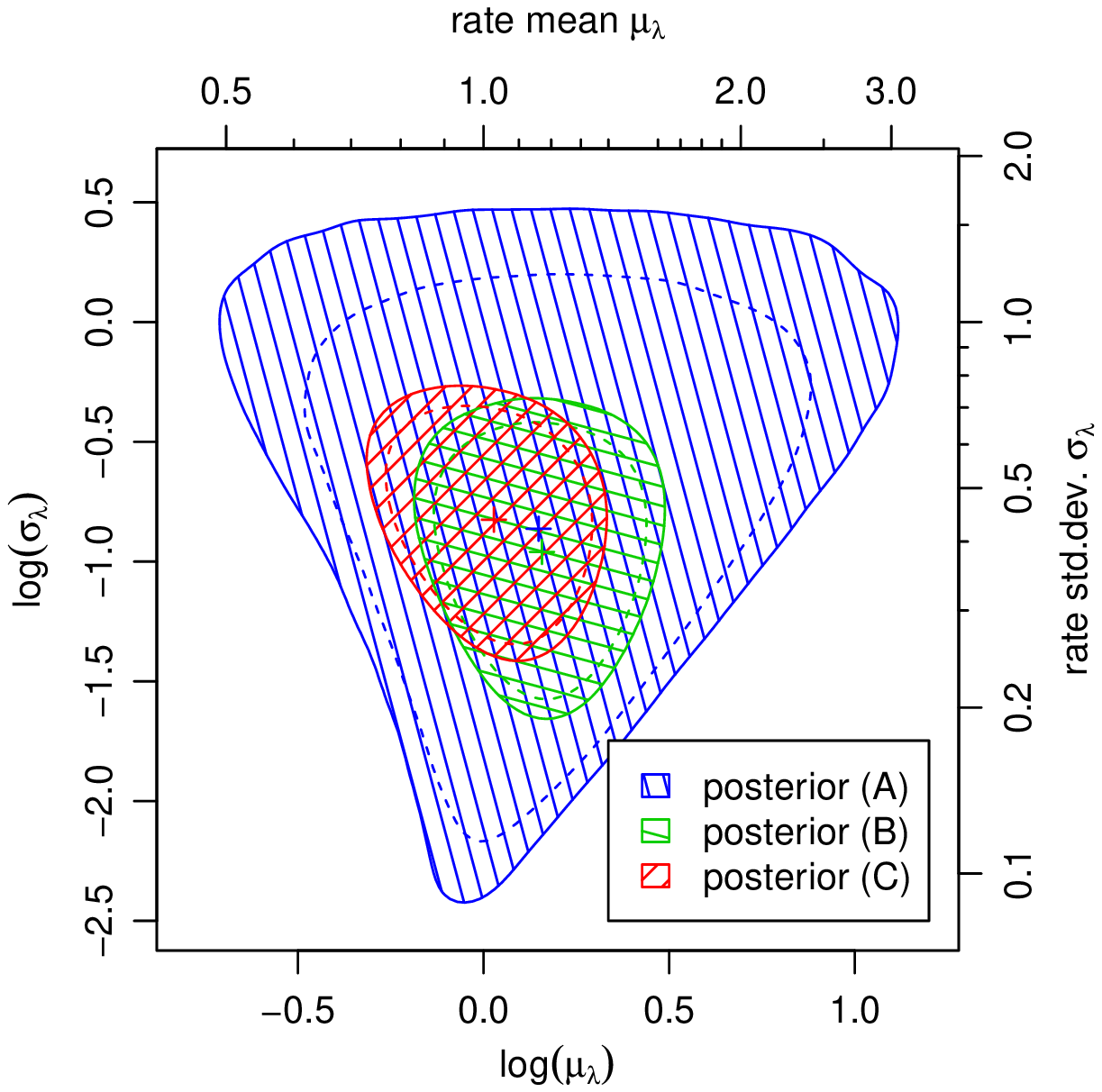}
      \hspace{0.03\linewidth}
      \includegraphics[width=0.30\linewidth]{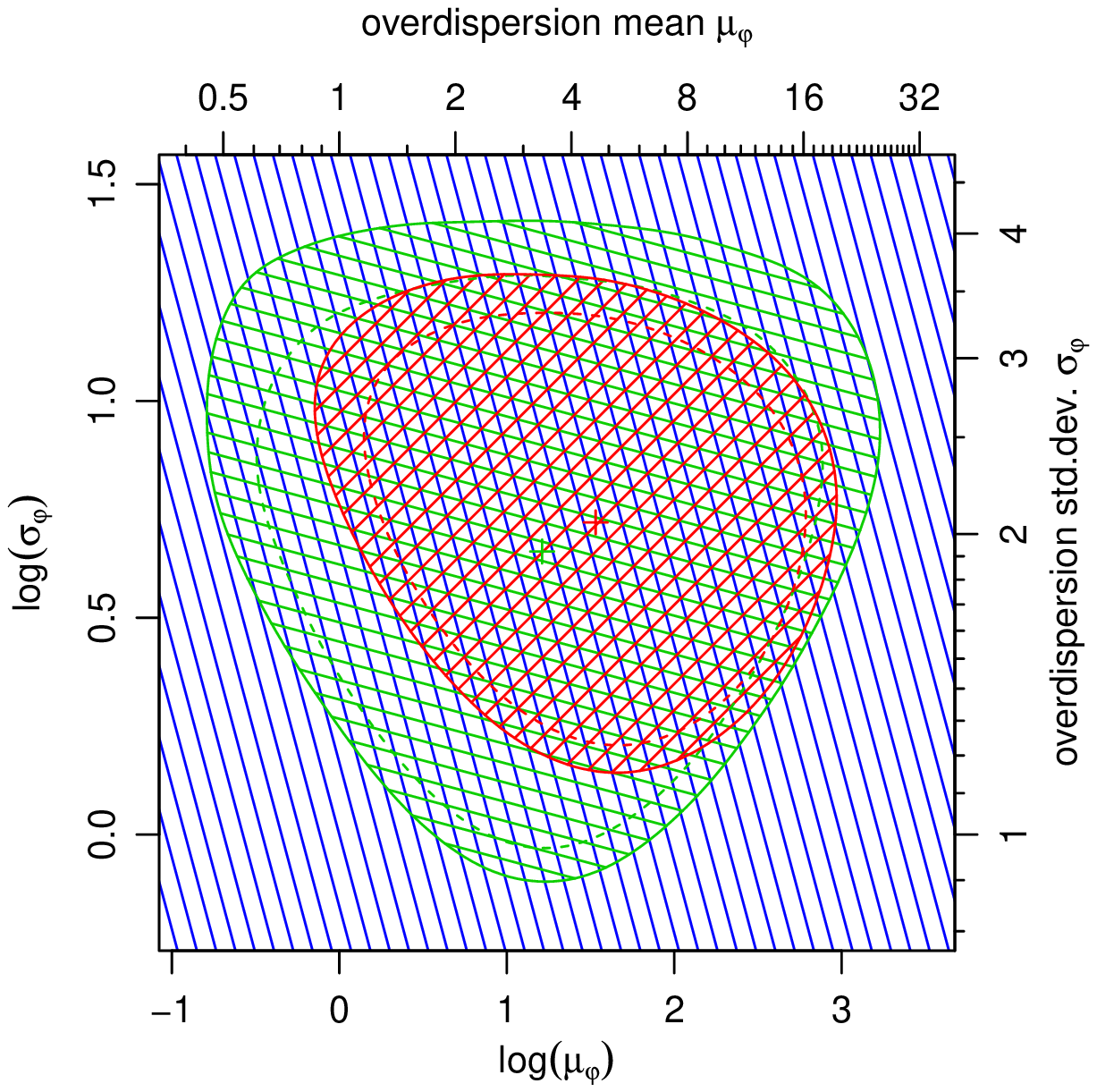}
      \hspace{0.03\linewidth}
      \includegraphics[width=0.30\linewidth]{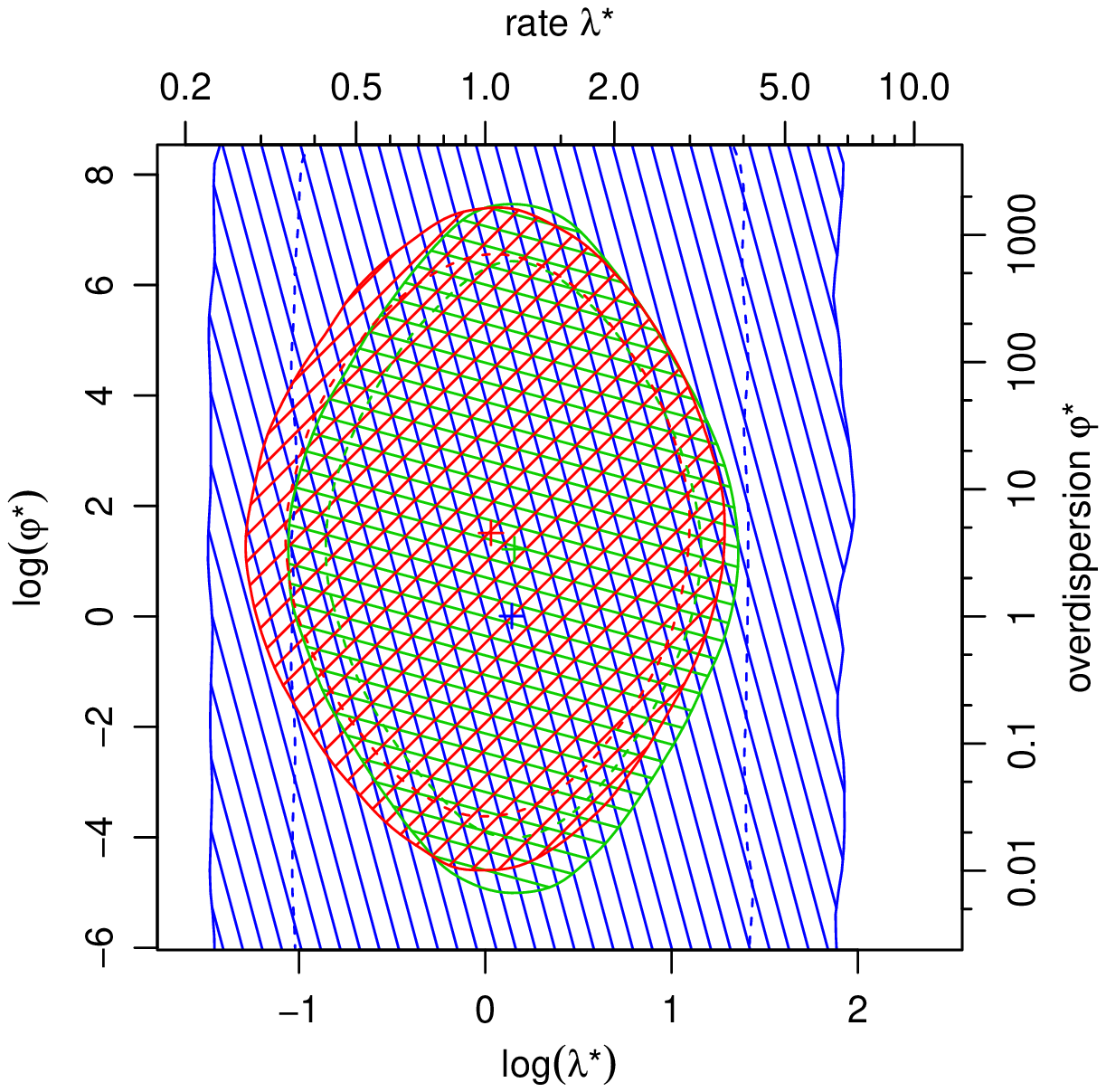}
      \caption{\label{fig:hyperParams}The left two panels show
        credibility areas for the four ``population'' rate and
        overdispersion hyperparameters $(\mu_\lambda,\sigma_\lambda)$
        and $(\mu_\varphi,\sigma_\varphi)$.  Solid lines enclose
        95\%, and dashed lines 90\% probability; the crosses indicate
        median values.  The right panel illustrates the resulting
        posterior predictive distribution for the parameters in a
        ``new'' study $(\lambda^\ast,\varphi^\ast)$.}
    \end{center}
  \end{figure}

  Figure~\ref{fig:hyperParams} illustrates the information we gain
  about the rates and the amounts of overdispersion encountered in the
  analyzed studies.
  Again, considering different subsets of the studies, we gain more
  certainty about the four hyperparameters
  $(\mu_\lambda,\sigma_\lambda)$ and $(\mu_\varphi,\sigma_\varphi)$
  describing mean and variability in rate and overdispersion among
  different studies with the inclusion of additional studies.
  While the resulting posterior distributions
  differ, they are consistent and exhibit significant overlap. Most
  notably, in scenario~A 
  where we only consider studies that provide rate estimates along
  with standard errors, we do not gain any information on the
  overdispersion~$\varphi$, so that here the posterior equals the
  prior.  The right panel shows the posterior predictive distribution
  of rate and overdispersion~$(\lambda^\ast,\varphi^\ast)$ of a
  ``new'' study, illustrating what we have learned from the present
  set of studies about probable characteristics of an additional study
  from the same population; this would be of interest e.g.\ when
  planning a future study.

  We also investigated whether we could find any particular reason for
  the differences in effect estimates based on different data
  subsets. While the results do not look inconsistent, it is
  interesting that the analyses based on subsets (B) and (C) tend to
  indicate a stronger treatment effect. One notable feature of studies
  in subset~(A) is that these tend to be fairly long studies in
  comparison.  

\section{Discussion}
  In the context of meta-analysis, one is commonly faced with the
  problem of coherently reconciling data from different sources within
  a single model in order to infer quantities of interest.
  Here we were able to utilize data in the form of rate
  estimates and odds ratios, and, everything being based on an
  underlying negative binomial model, differing follow-up times are
  not an issue, overdispersion is addressed, and inference may be
  concentrated directly on the parameter of primary interest, the
  treatment effect on the event rate.  The use of a Bayesian framework
  easily allowed for a flexible specification of correlation
  structures, including random effects at different levels.
  The joint likelihood formulation here allowed for the coherent
  incorporation of information in terms of quoted event rates or
  proportions of event-free patients. This rendered those studies
  providing both figures most valuable, but also enabled the inclusion
  of a considerable number of studies providing only one of the
  two. The utilization of additional data sources improved parameter
  estimates and reinforced the results' validity.

  The approach may in future be extended further by providing an
  interface for data in terms of estimated survival times (e.g.,
  median time to first event), which would here need to be implemented
  via the corresponding distribution of survival times (Exponential
  distribution in case of a Poisson, and a Lomax distribution in case
  of a negative binomial model \citep{SiriEtAl2012}).  While by now we
  incorporated studies providing rates and standard error via a
  conventional normal approximation to the likelihood, this could as
  well be implemented via another (approximate) negative binomial
  likelihood, e.g., by assuming that standard errors are based on the
  emprical standard deviation, in which case this figure again carries
  information on the amount of overdispersion.  For the moment we
  assumed independence between rate and overdispersion parameters, but
  introducing correlation between the two may also constitute 
  a plausible and useful extension, adding robustness to the approach.
  Bivariate meta-analyses of the rate and overdispersion estimates for placebo control groups and active treatments, however, do not indicate the presence of correlation between the two parameters in our example data.
  Similarly, one may also question the
  assumption of equal overdispersion across arms within a single study.
  For our example, we checked the assumption by performing a meta-analysis based on placebo- and treatment-arm overdispersions, which does not provide evidence against the assumption of a common overdispersion parameter across groups.
  However, the chances of resolving such issues based on the given
  data may be small, such investigations may be more realistically
  addressed based on individual patient data.  Another further
  extension of the presented method would be to utilize information on
  potential missingness of data that may be deducible from the given
  set of
  studies \citep[e.g.][]{DuvalTweedie2000,Copas2013,CopasEtAl2014}.
  Up to now the type of measure reported (rate and/or zero counts) is
  assumed to be independent of the particular trial's outcome. We have
  already seen that the way of reporting may be connected with other
  features like study duration (see Sec.~\ref{sec:results}). It would
  be interesting to consider additional information (e.g. from study
  protocols) in order to test for potential correlations and biases
  here. The general problem remains to meet the two-fold challenge of
  first \textsl{finding} all relevant data and then
  also \textsl{utilizing} it in a coherent mannner.  Modeling would of
  course be much simplified if data were available not in the form of
  study-level summary statistics, but in terms of detailed
  individual-patient data. While the fully Bayesian approach using
  MCMC methods yields accurate results, it is computationally rather
  costly, so it may be worth investigating alternative approaches
  (like empirical Bayes quasi-likelihood methods) in comparison.
  It would have been of interest to validate the approach with
  external individual patient data, however, we did not have access to
  this kind of data and cannot pursue this line of investigation. As
  suggested by a referee, a leave-one-out cross validation would be an
  excellent alternative validation, however, due to the associated
  computational costs, this is currently not feasible; we will further
  pursue this matter, possibly also by implementing checks via
  predictive distributions, along with the investigation of
  computational speed-ups.
  It should also be relatively easy to carry over the approximations
  used here to related models, for example to other generalizations of
  the Poisson distribution like the zero-inflated Poisson
  distribution.

\subsubsection*{Acknowledgement}
  Part of the work was funded by a unrestricted contribution (Oskar
  und Helene Medizinpreis 2010) to SA.

\subsubsection*{Conflict of interest}
{The authors have declared no conflict of interest.}

\section*{Appendix}
\subsection*{A.1 Rate ratios and odds ratios}\label{sec:RateOrAppendix}
  Consider the odds for the occurrence of at least one event in the
  Poisson model.  We have: $\prob(X\!=\!0)=\exp(-d\lambda)$ and
  $\prob(X\!\geq\! 1)=1-\exp(-\delta\lambda)$.  Assuming a multiplicative
  treatment effect $\vartheta<1$ on the rate, the odds ratio is
  \begin{eqnarray}
    \frac{1-\exp(-\delta\vartheta\lambda)}{\exp(-\delta\vartheta\lambda)} \times \frac{\exp(-\delta\lambda)}{1-\exp(-\delta\lambda)}
    &=& 
    \frac{\exp(\delta\vartheta\lambda)-1}{\exp(\delta\lambda)-1}
  \end{eqnarray}
  which is $\approx \vartheta$ for small $\delta\lambda$, and smaller
  otherwise, i.e., the effect on odds ratio is larger than the effect
  on rate, and the effect will appear more pronounced for longer
  studies.

  In the negative binomial model, we have: $\prob(X\!=\!0)=(1+\varphi
  \delta \lambda)^{-\frac{1}{\varphi}}$ and $\prob(X\!\geq\! 1)=1-(1+\varphi
  \delta \lambda)^{-\frac{1}{\varphi}}$.  Again assuming a treatment effect
  $\vartheta<1$, the odds ratio is
  \begin{eqnarray}
    \frac{1-(1+\varphi \delta \vartheta\lambda)^{-\frac{1}{\varphi}}}{(1+\varphi \delta \vartheta\lambda)^{-\frac{1}{\varphi}}} \times \frac{(1+\varphi \delta \lambda)^{-\frac{1}{\varphi}}}{1-(1+\varphi \delta \lambda)^{-\frac{1}{\varphi}}}
    &=&
    \frac{(1+\varphi \delta \vartheta \lambda)^\frac{1}{\varphi}-1}{(1+\varphi \delta \lambda)^\frac{1}{\varphi}-1}
  \end{eqnarray}
  which is $\approx \vartheta$ for small $\delta\lambda$, $<\vartheta$ for
  $\varphi<1$, $=\vartheta$ for $\varphi=1$ and $>\vartheta$ for
  $\varphi>1$, i.e., the effect on the odds ratio again is
  approximately the same as the effect on $\theta$ for small event
  rate~$\lambda$ or exposure duration~$\delta$, and whether the effect in
  terms of odds ratio is smaller or greater than in terms
  of~$\vartheta$ otherwise depends on the amount of
  overdispersion~$\varphi$.

\subsection*{A.2 Marginal and joint distributions of total and zero counts}\label{sec:JointEstimationAppendix}
  Suppose we have negative binomial random variables $X_1,\ldots,X_n$
  with rate~$\lambda$, overdispersion~$\varphi$ and a common
  corresponding exposure duration~$\delta$.  The (approximate) marginal
  distributions of the total count~$T$ and the number of
  zero-counts~$Z$ are described in Section~\ref{sec:TZcounts}.
  The \textsl{joint} distribution of $T$ and $Z$ may be derived by
  noting that the density function may be factorised as
  $p(t,z)=p(z) \times p(t|z)$.  The \textsl{conditional} distribution
  of $T|Z$ is essentially that of a sum of $n\!-\!z$ draws from a
  ``truncated'' negative binomial distribution only taking positive
  values, i.e., the conditional distribution of a negative binomial
  draw~$X$ given that $X\!\neq\! 0$.  Instead of the moments given in
  (\ref{eqn:NBMoments}), the ``truncated'' distribution instead has
  the expectation
  \begin{equation}
    \expect[X|X\!>\!0] 
    \;=\;  {\textstyle\frac{1}{1-\prob(X\!=\!0)}} \sum_{j=1}^\infty j\times\prob(X\!=\!j)
    \;=\; {\textstyle\frac{\expect[X]}{1-\prob(X\!=\!0)}} \nonumber
  \end{equation}
  which for the general negative binomial case yields:
  \begin{equation}
    \expect[X|X\!>\!0]
     \;=\; 
    \left\{\begin{array}{ll}
          {\textstyle \frac{\delta \lambda}{1-(\frac{1}{1+\varphi \delta \lambda})^{1/\varphi}}} & \mbox{for } \varphi>0 \\
          {\textstyle \frac{\delta \lambda}{1-\exp(-\delta\lambda)}} & \mbox{for } \varphi=0.
          \end{array}\right.\nonumber
  \end{equation}
  The variance is 
  \begin{eqnarray}
    \var(X|X\!>\!0) 
    &=& \sum_{j=1}^\infty \bigl(j-\expect[X|X\!>\!0]\bigr)^2 \times \prob(X\!=\!j|X\!>\!0) \nonumber
    \\
    &=& \bigl(\expect[X] - \expect[X|X\!>\!0]\bigr)^2 + \sum_{j=1}^\infty \bigl(j-\expect[X]\bigr)^2 \, \prob(X\!=\!j|X\!>\!0)\nonumber
    \\
    &=& \bigl(\expect[X] - \expect[X|X\!>\!0]\bigr)^2 + \frac{1}{1-\prob(X\!=\!0)} \sum_{j=1}^\infty \bigl(j-\expect[X]\bigr)^2 \, \prob(X\!=\!j)\nonumber
    \\
    &=& \bigl(\expect[X] - \expect[X|X\!>\!0]\bigr)^2 \nonumber\\
     && + \frac{1}{1-\prob(X\!=\!0)} \Bigl\{ \sum_{j=0}^\infty \bigl(j-\expect[X]\bigr)^2 \, \prob(X\!=\!j)  - \expect[X]^2\, \prob(X\!=\!0)
        \Bigr\}\nonumber
    \\
    &=& \frac{\var(X) - \expect[X]^2\, \prob(X\!=\!0)}{1-\prob(X\!=\!0)} + \bigl(\expect[X] - \expect[X|X\!>\!0]\bigr)^2 \nonumber
    \\
    &=& \frac{\var(X) - \expect[X]^2\, \prob(X\!=\!0)}{\prob(X\!>\!0)} + \expect[X]^2 \Bigl(\frac{\prob(X\!=\!0)}{\prob(X\!>\!0)}\Bigr)^2 \nonumber
  \end{eqnarray}
  which for the negative binomial case ($\varphi>0$) yields
  \begin{equation}
    \var(X|X\!>\!0) 
    \;=\; 
          \frac{\delta\lambda}{1-(1+\varphi \delta \lambda)^{-\frac{1}{\varphi}}}
          +(\delta\lambda)^2\frac{2(1+\varphi \delta \lambda)^{-\frac{2}{\varphi}} - (1+\varphi)(1+\varphi \delta \lambda)^{-\frac{1}{\varphi}} + \varphi}{\bigl(1-(1+\varphi \delta \lambda)^{-\frac{1}{\varphi}}\bigr)^2}\nonumber\label{eqn:NbCondVar}
  \end{equation}
  and for the Poisson case ($\varphi=0$)
  \begin{equation}
    \var(X|X\!>\!0) 
    \;=\; 
          \delta\lambda + \frac{(\delta\lambda)^2}{\bigl(\exp(\delta\lambda)-1\bigr)^2} + \frac{\delta\lambda-(\delta\lambda)^2}{\exp(\delta\lambda)-1}.\nonumber\label{eqn:PoiCondVar}
  \end{equation}
  Again assuming a normal approximation we then get an (approximate) joint density of $T$ and $Z$
  \begin{eqnarray}
    p(t, z) &=& p(z) \times p(t|z) \nonumber
    \\
    &=& {n \choose z} \, \pi_0^z \, 
        (1-\pi_0)^{n-z} 
        \times \frac{1}{\sqrt{2\pi(n\!-\!z)\sigma^2}} \,\exp\Bigl(\textstyle -\frac{1}{2}\frac{(t -(n\!-\!z)\theta)^2}{(n\!-\!z)\,\sigma^2}\Bigr)\nonumber
  \end{eqnarray}
  where
  \begin{eqnarray}
    \pi_0      &=&
    \left\{\begin{array}{ll}
          {\textstyle (1+\varphi \delta \lambda)^{-\frac{1}{\varphi}}} & \mbox{for } \varphi>0 \\
          {\textstyle \exp(-\delta\lambda)} & \mbox{for } \varphi=0
          \end{array}\right.\nonumber\\
    \theta   &=&
          {\textstyle\frac{\delta \lambda}{1-\pi_0}} \nonumber\\
    \sigma^2 &=&
    \left\{\begin{array}{ll}
            \theta + (\delta\lambda)^2\,\frac{2\pi_0^2 - (1+\varphi)\pi_0 + \varphi}{(1-\pi_0)^2} & \mbox{for } \varphi>0 \\
            \delta\lambda + \frac{(\delta\lambda)^2}{(\frac{1}{\pi_0}-1)^2} + \frac{\delta\lambda-(d\lambda)^2}{\frac{1}{\pi_0}-1}    & \mbox{for } \varphi=0 . \nonumber
          \end{array}\right.
  \end{eqnarray}